\documentclass{PoS}
\usepackage{amsmath}
\usepackage{graphicx}
\usepackage{epstopdf}
\def\procspie{\ref@jnl{Proc.~SPIE}}   

\usepackage{soul}
\usepackage{float}
\usepackage{fixltx2e}

\title{Performance of the MAGIC telescopes under moonlight}

\ShortTitle{Performance of the MAGIC telescopes under moonlight}

\author{\speaker{D. Guberman}$^{1}$ and P. Colin$^{2}$ for the MAGIC collaboration\thanks{We would like to thank the IAC for the excellent working conditions at the ORM in La Palma. We acknowledge the financial support of the German BMBF, DFG and MPG, the Italian INFN and INAF, the Swiss National Fund SNF, the European ERDF, the Spanish MINECO, the Japanese JSPS and MEXT, the Croatian CSF, and the Polish MNiSzW.}\\
       $^{1}$Institut de Fisica d'Altes Energies (IFAE), The Barcelona Institute of Science and Technology, Campus UAB, 08193 Bellaterra (Barcelona), Spain\\
       $^{2}$ Max-Planck-Institut f\"ur Physik, D-80805 M\"unchen, Germany \\
        E-mail: \email{dguberman@ifae.es}}

\abstract{MAGIC, a system of two imaging atmospheric Cherenkov telescopes, achieves its best performance under dark conditions, i.e. in absence of moonlight or twilight. Since operating the telescopes only during dark time would severely limit the duty cycle, observations are also performed when the Moon is present in the sky. Here we present a dedicated Moon-adapted analysis and characterize the performance of MAGIC under moonlight. We evaluate energy threshold, angular resolution and sensitivity of MAGIC under different background light levels, based on Crab Nebula observations and tuned Monte Carlo simulations. This study includes observations taken under non-standard hardware configurations, such as reducing the camera photomultiplier tubes gain by a factor $\sim$1.7 (reduced HV settings) with respect to standard settings (nominal HV) or using UV-pass filters to strongly reduce the amount of moonlight reaching the telescopes cameras. The Crab Nebula spectrum is correctly reconstructed in all the studied illumination levels, that reach up to 30 times brighter than under dark conditions. The main effect of moonlight is an increase in the analysis energy threshold and in the systematic uncertainties on the flux normalization. The sensitivity degradation is constrained to be below 10\%, within 15-30\% and between 60 and 80\% for nominal HV, reduced HV and UV-pass filter observations, respectively. No worsening of the angular resolution was found. Thanks to observations during moonlight, the duty cycle can be doubled, suppressing the need to stop observations around full Moon.}

\FullConference{35th International Cosmic Ray Conference -- ICRC217-\\
		10-20 July, 2017\\
		Bexco, Busan, Korea}

\begin{document}

\section{Introduction}\label{sec:Intro}

The Imaging Atmospheric Cherenkov Technique (IACT) uses one or several optical telescopes to image the air showers induced by cosmic $\gamma$ rays in the atmosphere. IACT telescope arrays usually achieve their best performance in the absence of moonlight. Their cameras are generally equipped with photomultiplier tubes, that can age (gain degradation with time) quickly in a too bright environment. As a result, observations are restricted to relatively dark conditions. When IACT instruments operate only during moonless astronomical nights, their duty cycle is limited to 18\% ($\sim$1500\,h/year), without including the observation time loss due to bad weather or technical issues. Every month around the full Moon, the observations are generally fully stopped for several nights in a row. Operating IACT telescopes during moonlight and twilight time would allow increasing the duty cycle up to $\sim$40\%.

MAGIC (Major Atmospheric Gamma-ray Imaging Cherenkov telescopes) consists of two 17~m diameter IACT telescopes designed to observe very high energy (VHE, $\gtrsim$ 50 GeV) $\gamma$-rays\cite{upgrade1}.The cameras of the MAGIC telescopes were designed from the beginning to allow observations during moderate moonlight \cite{OldMoon, MagicMoon-ICRC2009}. This is achieved by operating the PMTs at a relatively low gain, typically of 3-4~$ \times 10^4$. The use of reduced HV \cite{MagicMoonShadow} and UV-pass filters \cite{Filters} were introduced later to extend the observations to all the possible Night Sky Background (NSB) levels, up to few degrees from a full Moon.

IACT observations under moonlight are becoming more and more standard, and
are routinely performed with the MAGIC and VERITAS\cite{VERITAS2008} telescopes. The
performance of VERITAS under moonlight with different hardware settings at
a given NSB level has been recently reported \cite{VeritasNew}. Here we present an overview of the results reported in \cite{MAGIC_moon}: a more complete study on how the performance of an IACT instrument is affected by moonlight and how it degrades as a function of
the NSB. Our study is based on extensive observations of the Crab Nebula,
adapted data reduction and tuned Monte Carlo (MC) simulations. The
observations, carried out from October 2013 to March 2016 by MAGIC with
nominal HV, reduced HV and UV-pass filters, cover the full range of NSB
levels that are typically encountered during moonlight nights.

\section{The MAGIC telescopes under moonlight}\label{sec:MAGIC}

Each camera of the MAGIC telescopes consist of 1039 6-dynode PMTs. The aging of PMTs is determined by the amount of charge that hits the last dynode (anode). The higher the intensity of the light they are exposed to, the higher the collected charge. To avoid fast aging, observations  using the standard HV settings (nominal HV) are possible up to a brightness of about 12\,$\times \,\textit{NSB}_{\text{Dark}}$\footnote{$\textit{NSB}_{\text{Dark}}$ is defined in \cite{MAGIC_moon} as the brightness of the fraction of the sky seen by MAGIC when pointing the telescopes towards the Crab Nebula at low zenith angle during astronomical night, with no Moon in the sky or near the horizon, and good weather conditions.}. Observations can be extended up to about 20\,$\times  \, \textit{NSB}_{\text{Dark}}$ by reducing the gain of the PMTs by a factor $\sim$1.7 (reduced HV settings). When the HV is reduced there is less amplification in the dynodes and so fewer electrons hit the anode. However, the PMT gains cannot be reduced by an arbitrary large factor because the performance would significantly degrade, resulting in lower collection efficiency, slower time response, larger pulse-to-pulse gain fluctuations and an intrinsically worse signal-to-noise ratio \cite{MAGIC_moon, Photonis}.

Even when operating the telescopes with reduced HV, observations are severely limited or cannot be performed if the Moon phase is above $90\%$. Observations can, however, be extended up to about 100\,$\times  \, \textit{NSB}_{\text{Dark}}$ with the use of UV-pass filters. This limit is achievable if the filters are installed and at the same time PMTs are operated with reduced HV. This is done only in extreme situations ($>$50\,$\times  \, \textit{NSB}_{\text{Dark}}$). All the UV-pass filter data included in this work were taken with nominal PMT gain. In practice, observations can be performed in conditions that are safe for the PMTs as close as a few degrees away from a full Moon. The telescopes can be pointed almost at any position in the sky, regardless the Moon phase, and, as a result, they can be operated continuously without full Moon breaks.

\section{Data analysis}


The performance of MAGIC under moonlight was characterized using 174 hours of Crab Nebula observations taken between October 2013 and January 2016, under NSB conditions going from 1 (dark) up to $30 \times  \, \textit{NSB}_{\text{Dark}}$\footnote{Observations are possible at higher illumination levels, but it is hard to get Crab data under such occasions. In fact, only on rare situations MAGIC targets are found under higher NSB levels than the ones analyzed in this work.}. Data taken correspond to zenith angles ranging between 5$^\circ$ and 50$^\circ$
and were divided into different samples according to their NSB level and the hardware settings in which observations were performed (nominal HV, reduced HV or UV-pass filters), as summarized in Table~\ref{tabNoise}.

\begin{table*}[t]
\centering
\begin{tabular}{| c | c | c | c | c | c |}
\hline
Sky Brightness & Hardware Settings & Time & Pedestal Distr & Cleaning Level factors & Size Cut \\
               &      &       &  mean / rms    &  $\text{Lvl}_1$ / $\text{Lvl}_2$  &       \\
               
[$\textit{NSB}_{\text{Dark}}$]  &  & h &  [phe]  & [phe]     &   [phe]  \\
\hline
   \hline

     1 (Dark) & nominal HV & 53.5 & 2.0 / 1.0 &  6.0 / 3.5 &  50  \\
     1-2 & nominal HV & 18.9 & 2.5 / 1.2 & 6.0 / 3.5 &  60 \\
     2-3 & nominal HV & 13.2 & 3.0 / 1.3 & 7.0 / 4.5 &  80 \\
     3-5 & nominal HV & 17.0 & 3.6 / 1.5 & 8.0 / 5.0 & 110 \\
     5-8 & nominal HV & 9.8 & 4.2 / 1.7 & 9.0 / 5.5 & 150 \\
        \hline

     5-8 & reduced HV & 10.8 & 4.8 / 2.0 & 11.0 / 7.0 & 135 \\
     8-12 & reduced HV & 13.3 & 5.8 / 2.3 & 13.0 / 8.0 & 170 \\
     12-18 & reduced HV & 19.4 & 6.6 / 2.6 & 14.0 / 9.0 & 220 \\
        \hline

     8-15 & UV-pass filters & 9.5 & 3.7 / 1.6 & 8.0 / 5.0 & 100  \\
     15-30 & UV-pass filters & 8.3 & 4.3 / 1.8 & 9.0 / 5.5 & 135\\
   \hline
 \end{tabular}
 \caption{Effective observation time and noise level of the Crab Nebula subsamples in each of the NSB/hardware bins. The adapted image cleaning levels and size cuts used for their analysis are also shown.}
 \label{tabNoise}
 \end{table*}
 

The data have been analyzed using the standard MAGIC Analysis and Reconstruction Software (MARS, \cite{TrueMARS}) following the standard analysis chain described in \cite{upgrade2}, besides some additional modifications that were implemented to account for the different observation conditions. These modifications can be summarized into a set of dedicated image cleaning levels, Monte Carlo (MC) simulations and size cuts.

The image cleaning is performed right after the data calibration to remove those pixels that contain only noise. During moonlight observations the background fluctuations are higher and the cleaning levels must be increased accordingly. Those levels were modified to ensure that the fraction of pedestal events that contain only noise and survive the image cleaning is lower than 10\% and optimized to get the lowest possible analysis threshold for every bin. For more details about how image cleaning works in MAGIC refer to \cite{upgrade2}. The optimized cleaning levels for each bin are shown in Table~\ref{tabNoise}.

MC simulations have mainly two functions in the MAGIC data analysis chain. A first sample (train sample) is used to build look-up tables and multivariate decision trees (random forest), which are employed for the energy and direction reconstruction and gamma/hadron separation. A second, independent sample (test sample) is used for the telescope response estimation during the source flux/spectrum reconstruction. 

We prepared MC samples adapted for every NSB/hardware bin. For nominal and reduced HV settings, we used the standard MAGIC MC simulation chain with additional noise to mimic the effect of moonlight (and reduced HV). 
In the case of the UV-pass filter observations, additional modifications on the simulation chain were implemented to include the filter transmission and the shadowing produced by the frame ribs.

As MC simulations were not produced from the trigger level, they do not reproduce by themselves the effect of moonlight on the trigger (see section 3.10 in \cite{upgrade1} and section 2.1 in \cite{MAGIC_moon} for details on the MAGIC trigger system). Instead, we applied cuts on the sum of charge of pixels surviving the image cleaning (image size) on each telescope. This size cut acts as a software threshold and it is optimized bin-wise as the minimal size for which the data and MC distributions are matching.
The used size cuts are given Table~\ref{tabNoise}.

\section{Performance}


\subsection*{Energy threshold}

\begin{figure}[t]
\centering
\includegraphics[width=0.49\columnwidth]{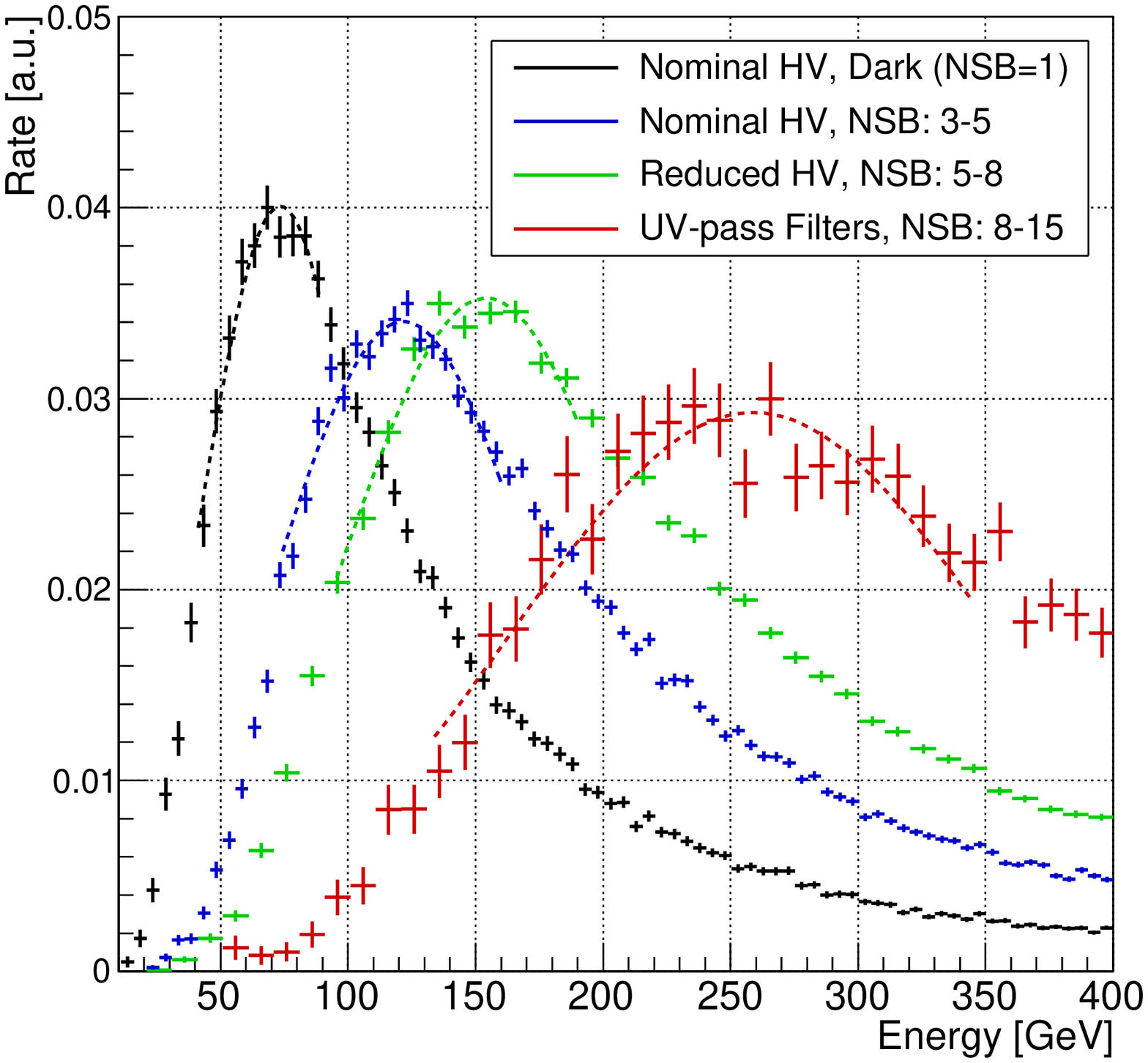}
\includegraphics[width=0.49\columnwidth]{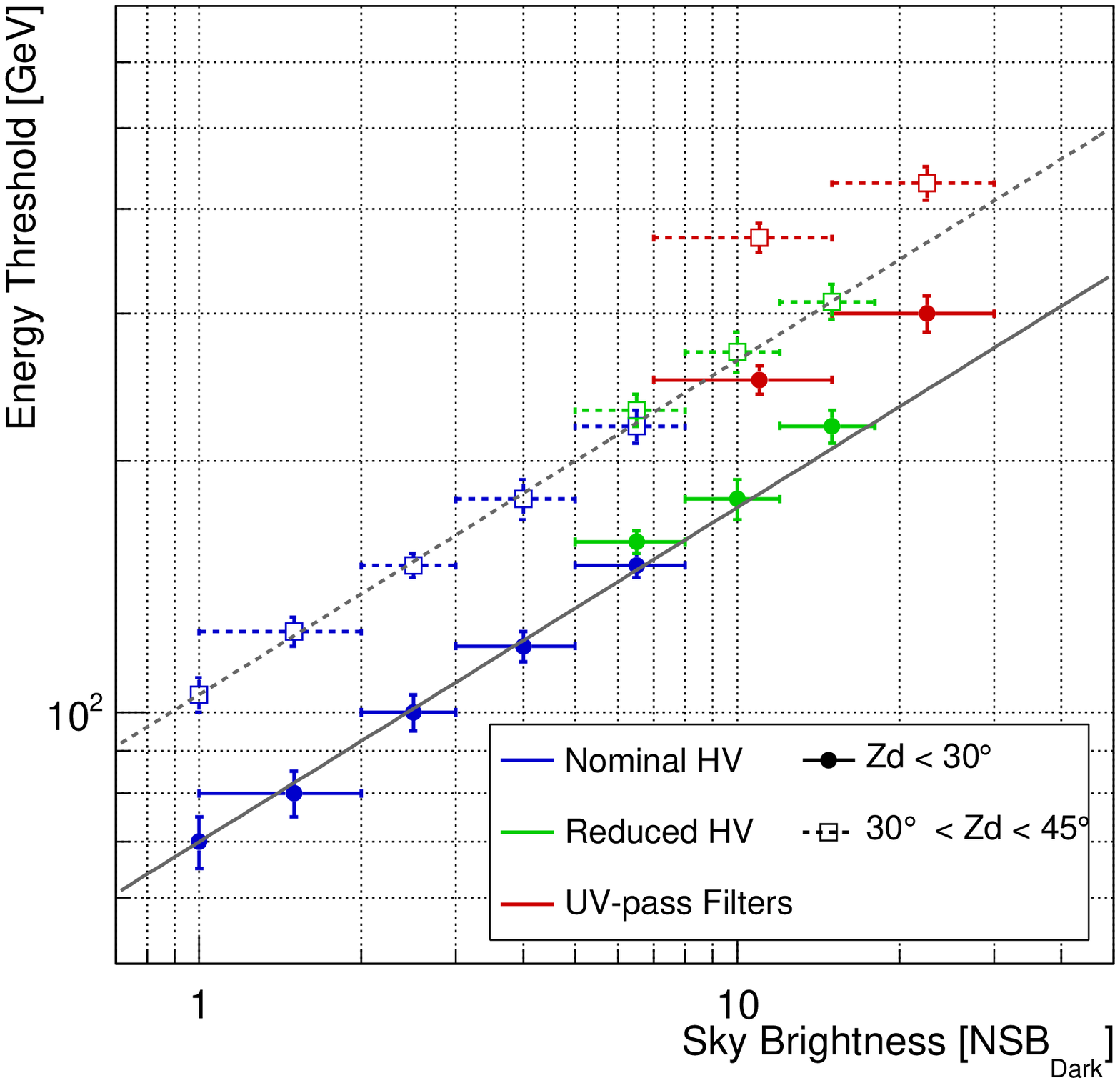}
\caption{\textbf{Left:} Rate of MC $\gamma$-ray events at the reconstruction  level for an hypothetical source with an spectral index of $-2.6$ observed at zenith angles below 30$^\circ$. 
Dashed lines show the gaussian fit applied to calculate the energy threshold on each sample. NSB levels are given in units of $\textit{NSB}_{\text{Dark}}$ \textbf{Right:} Energy threshold as a function of the sky brightness. 
The degradation of the energy threshold $E_{\text{th}}$ as a function of the NSB level can be roughly approximated, for nominal HV and reduced HV, by $E_{\text{th}}(\textit{NSB}) = E^{\text{Dark}}_{\text{th}} \times \left(\textit{NSB}/\textit{NSB}_{\text{Dark}}\right)^{0.4}$, where $E^{\text{Dark}}_{\text{th}}$ is the energy threshold during dark Crab Nebula observations.}\label{fig:EthNSB}
\end{figure}

The energy threshold of IACT telescopes is commonly defined as the peak of the differential event rate distribution as a function of energy. It is estimated from the effective collection area as a function of the energy, obtained from $\gamma$-ray MC simulations, multiplied by the expected $\gamma$-ray spectrum, which is typically (and also in this work) assumed to be a power-law with a spectral index of $-2.6$. Left panel in figure \ref{fig:EthNSB} shows four examples of such event rate distributions.

The energy threshold can be evaluated at different stages of the analysis. The lowest threshold corresponds to the trigger level, which reaches $\sim50$ GeV during MAGIC  observations in moonless nights at zenith angles below 30$^\circ$ \cite{upgrade2}.  Here we evaluate the energy threshold after image cleaning and size cuts are applied, for which a good matching between real data and MC is achieved. At this level, the previously quoted energy threshold rises to $\sim70$ GeV.

We produced differential rate plots for each NSB/hardware bin and estimated the energy threshold by fitting a Gaussian distribution in a narrow range around the peak of these distributions\footnote{Note that in those distributions the peak is broad, which means that it is possible to obtain scientific results with the telescopes below the defined threshold.}. Right panel in figure \ref{fig:EthNSB} shows the obtained energy threshold as a function of the sky brightness for different hardware configurations at low ($<30^\circ$) and medium ($30^\circ - 45^\circ$) zenith angles. 



\subsection*{Reconstruction of the Crab Nebula spectrum}

\begin{figure}[t]
\includegraphics[trim=0.5cm 0cm 1.5cm 0cm, clip=true, width=0.49\columnwidth]{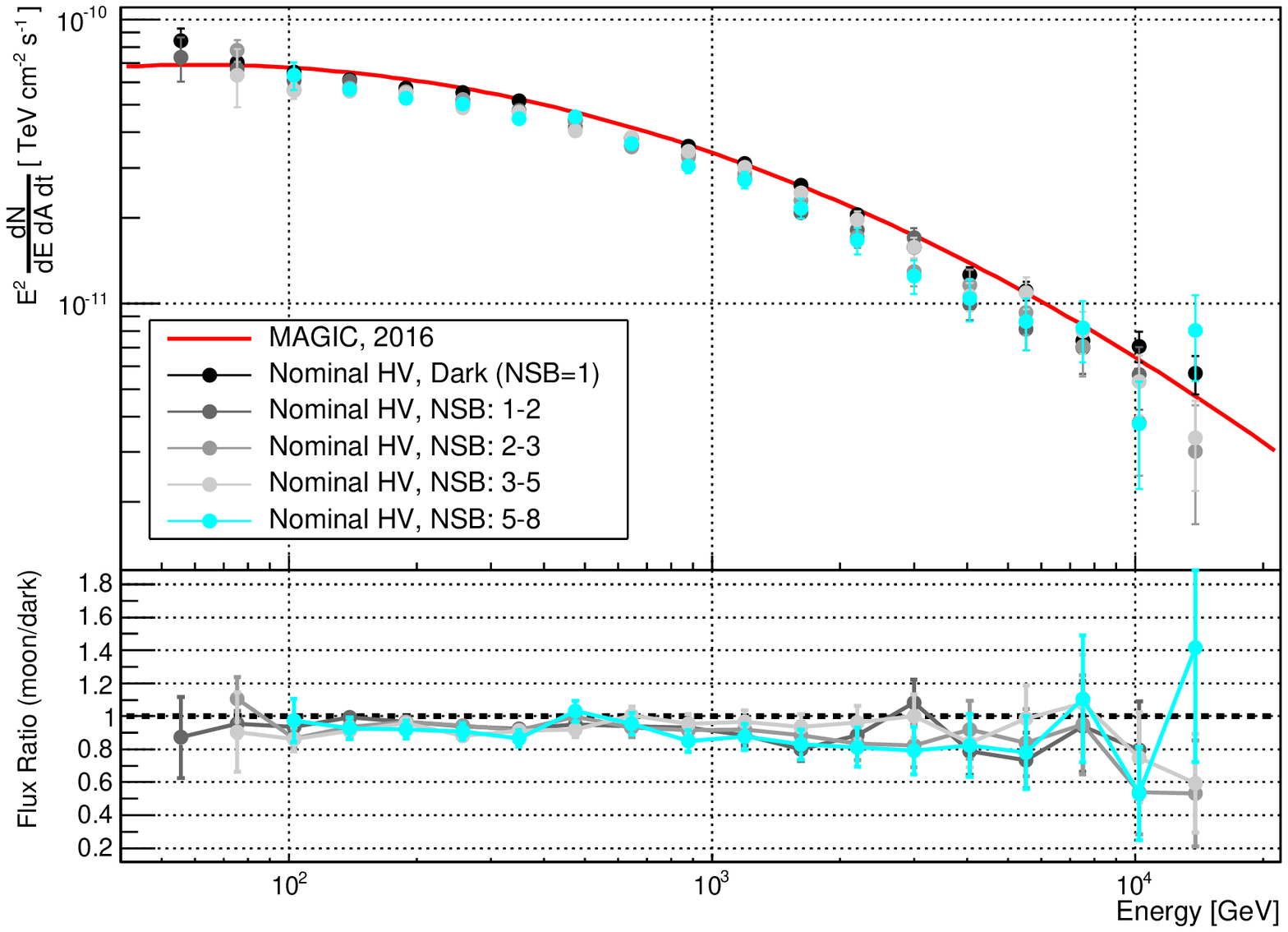}
\includegraphics[trim=0.5cm 0cm 1.5cm 0cm, clip=true, width=0.49\columnwidth]{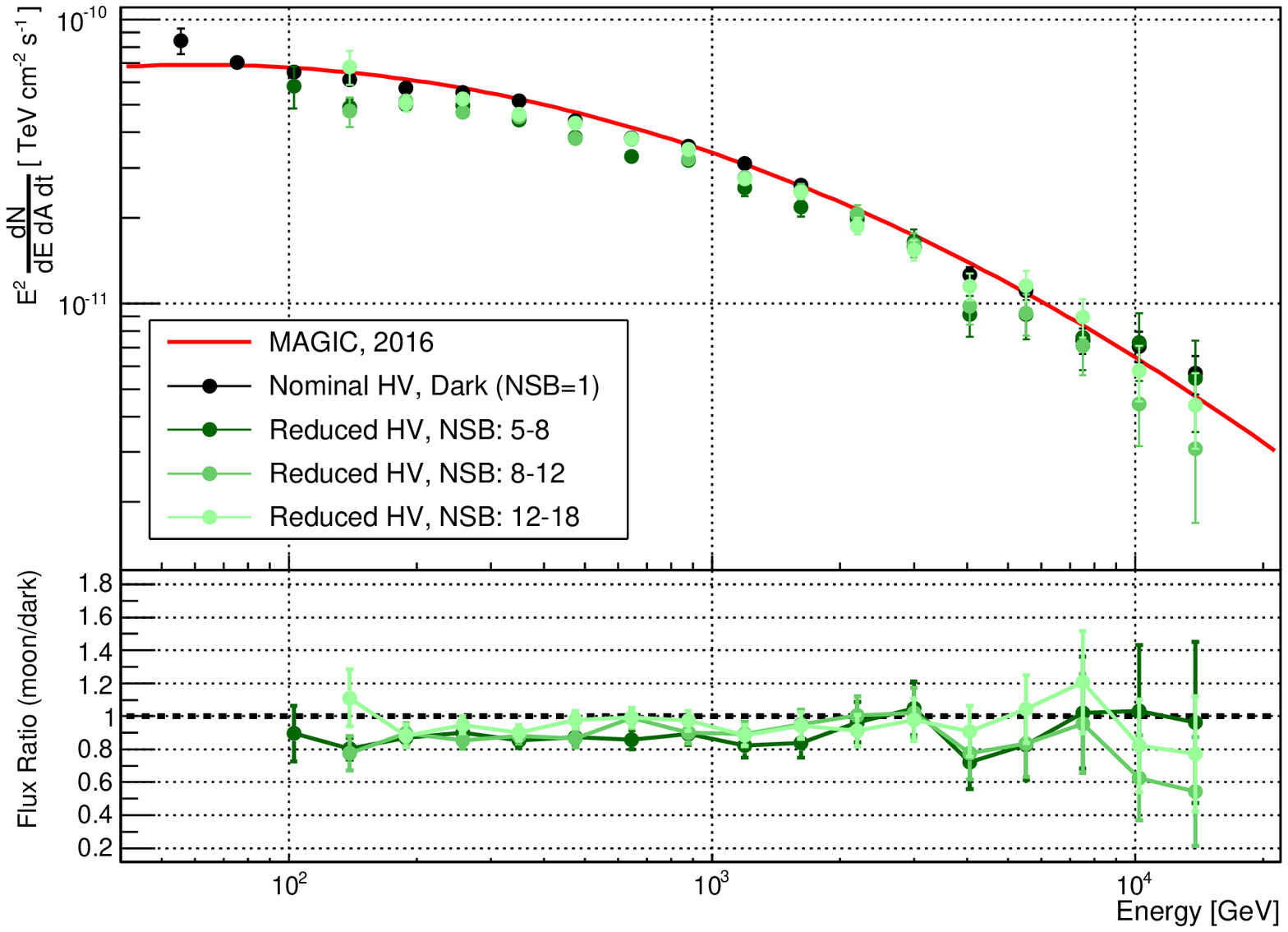}
\includegraphics[trim=0.5cm 0cm 1.5cm 0cm, clip=true, width=0.49\columnwidth]{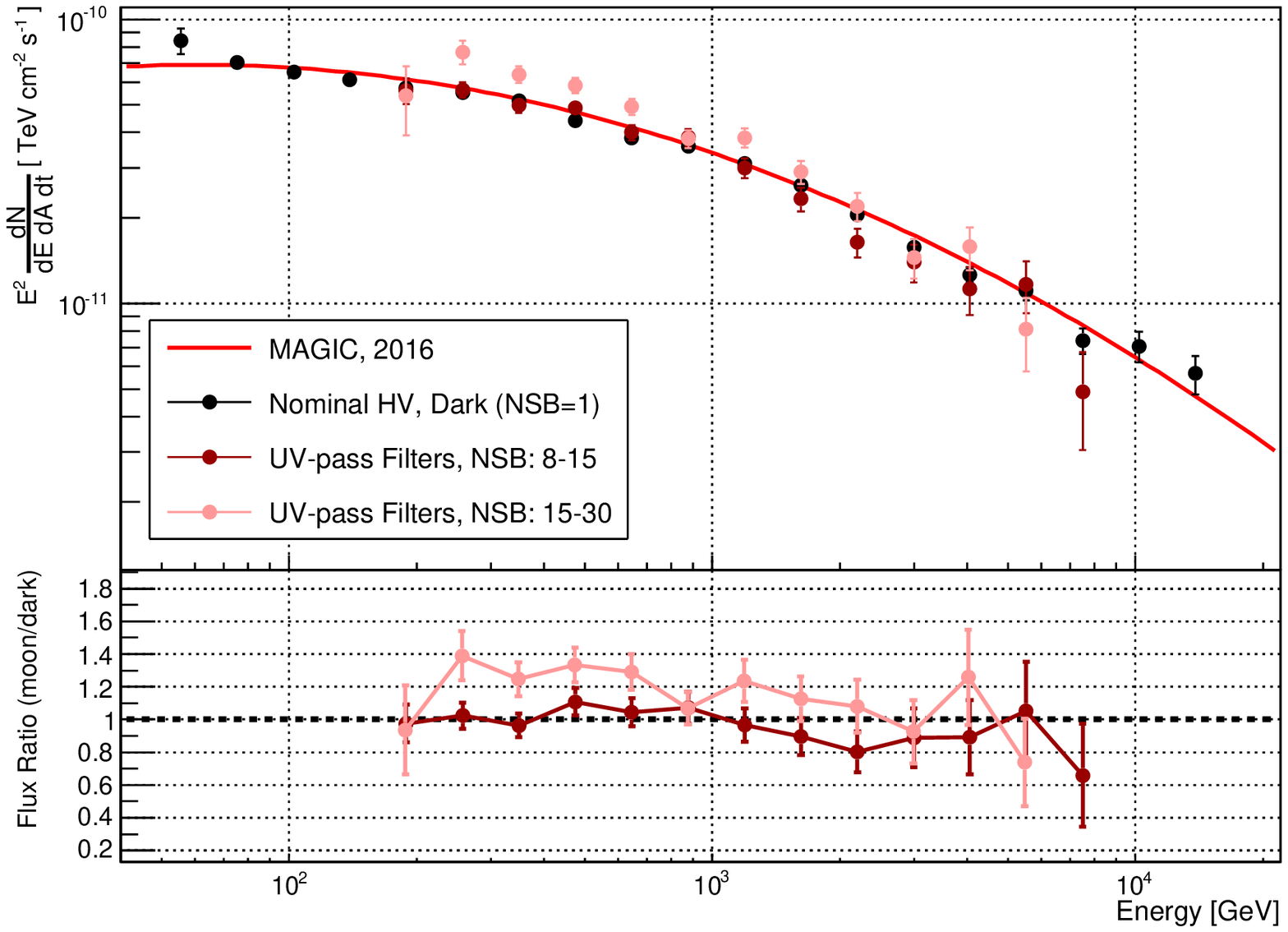}
\caption{Spectral energy distribution of the Crab Nebula obtained for different NSB levels (given in units of $\textit{NSB}_{\text{Dark}}$, coloured dots)  for nominal HV (top, left), reduced HV (top, right) and UV-pass filters (bottom, left) data. The results obtained with the dark sample using standard analysis in this work (black dots) and previously published by MAGIC (red solid line, \cite{upgrade2}) are shown in every panel. The bottom sub-panels show the ratio of the fluxes measured under moonlight to the flux measured under dark conditions.}\label{fig:CrabSEDMoon}
\end{figure}

The reconstructed Crab Nebula spectra obtained after applying the dedicated Moon analysis to each data set are shown in figure~\ref{fig:CrabSEDMoon}. In almost all the cases the fluxes obtained are consistent within $\pm$20\% with the one obtained under dark conditions, at least up to 4\,TeV. The only exception is the brightest NSB bin (UV-pass filters data up to 30 $\times \textit{NSB}_{\text{Dark}}$) where the ratio of the flux to the dark flux gets slightly above $\sim$30\% at energies between about 400 and 800\,GeV.

\subsection*{Sensitivity}

\begin{figure}
\includegraphics[width=0.5\columnwidth]{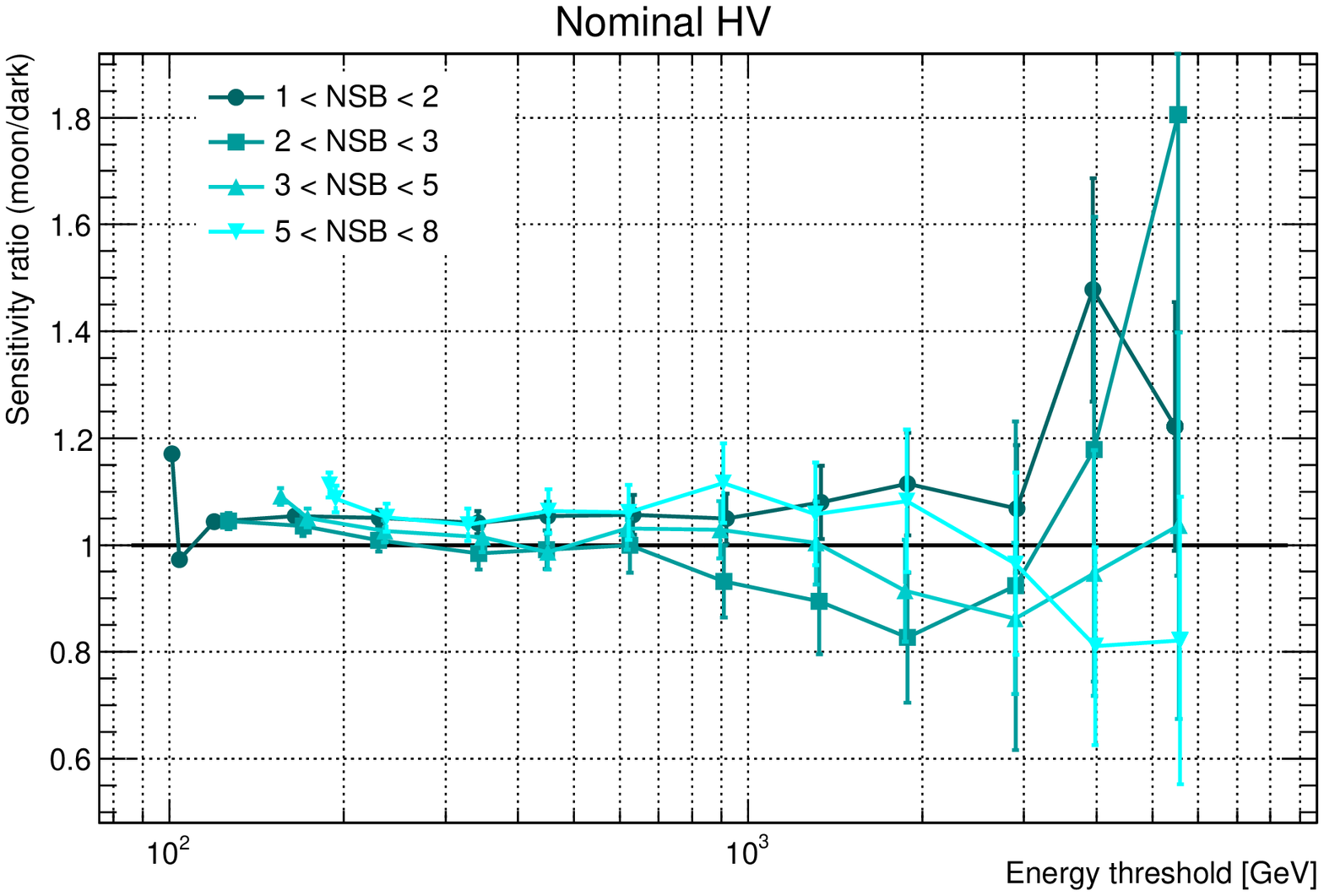}
\includegraphics[width=0.5\columnwidth]{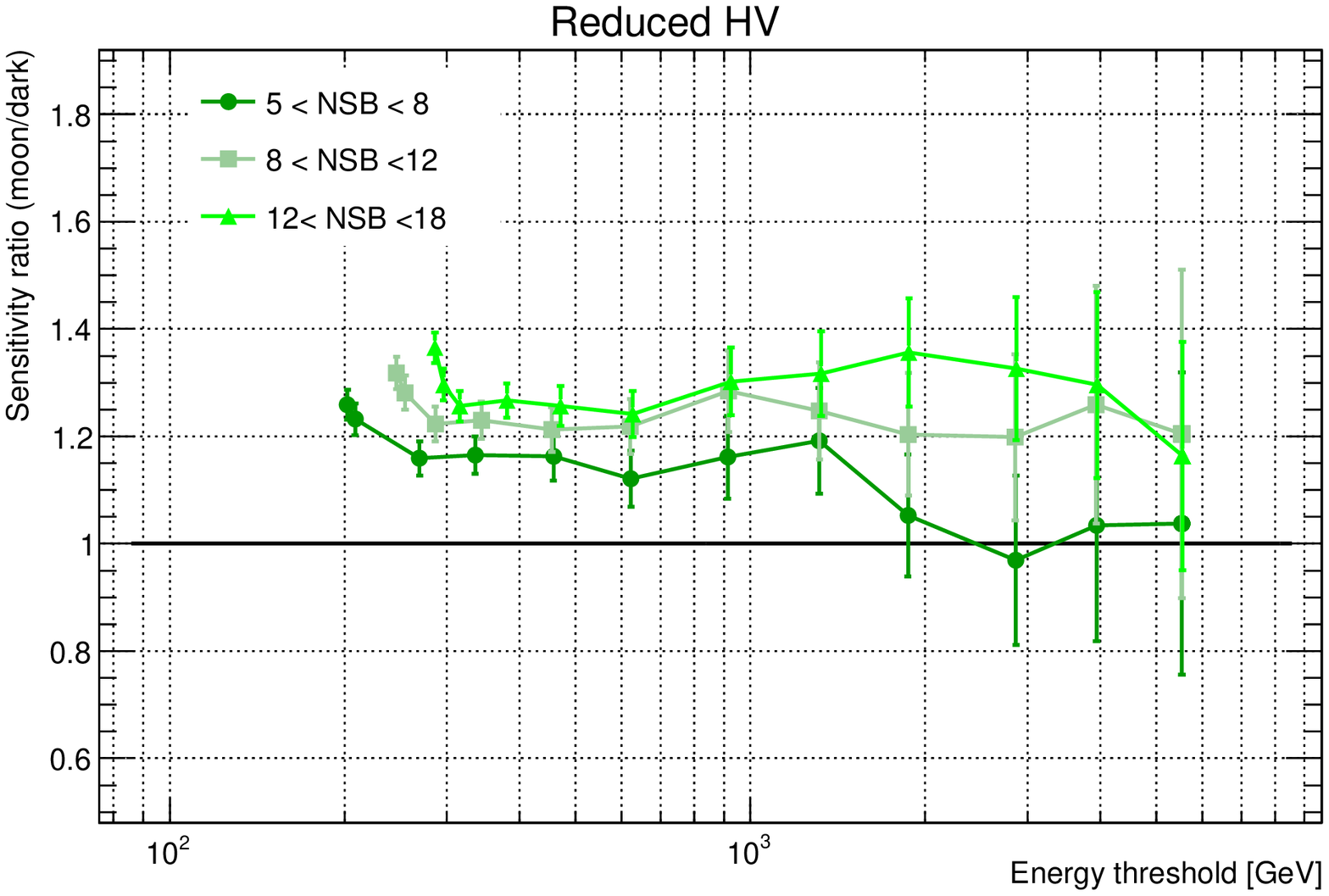}
\includegraphics[width=0.5\columnwidth]{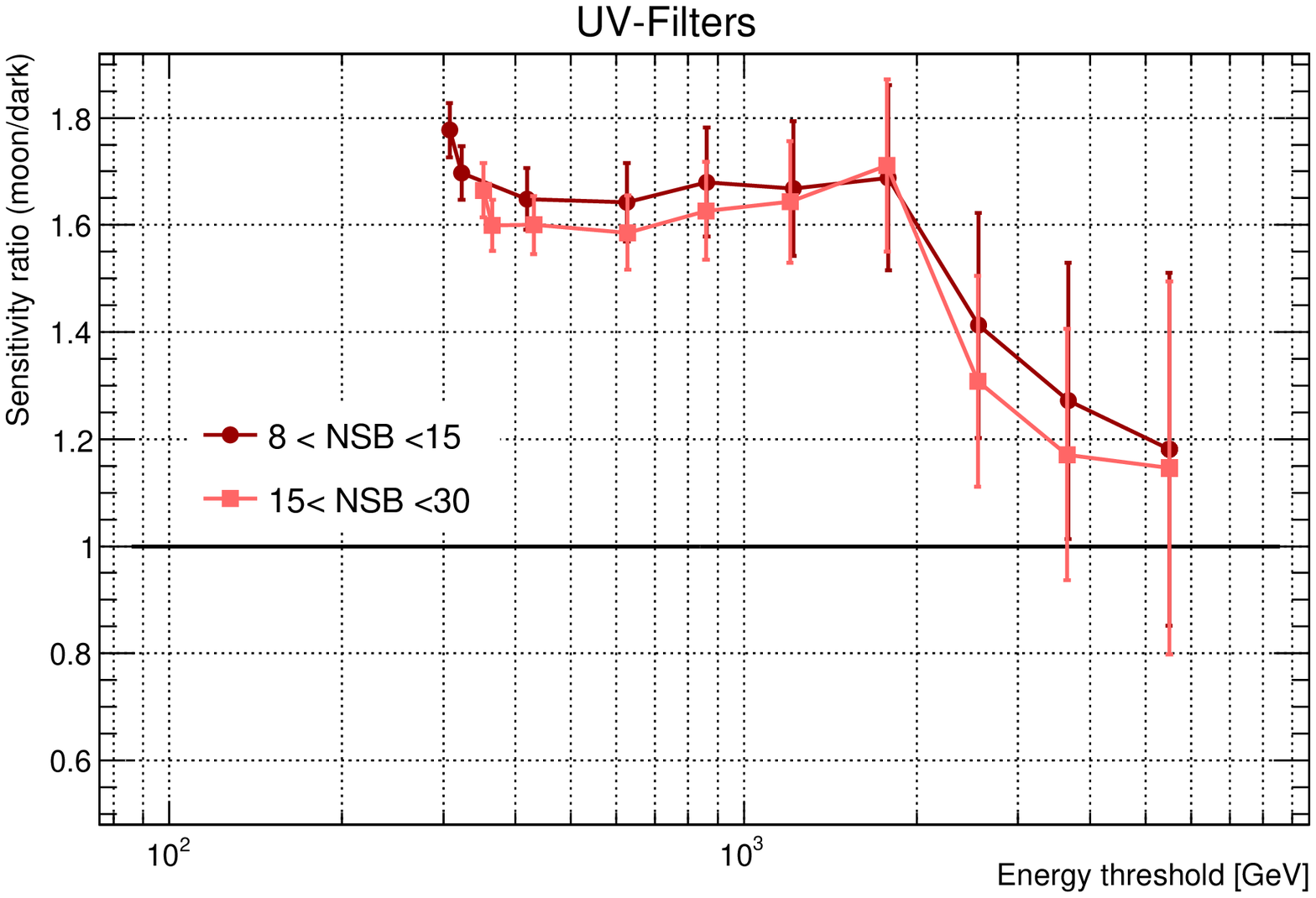}
\caption{Ratio of the integral sensitivity under moonlight to the dark sensitivity as a function of the analysis energy threshold. The NSB levels are given in unit of $\textit{NSB}_{\text{Dark}}$}\label{fig:Sens}
\end{figure}

Following the method described in \cite{upgrade2}, we evaluated the integral sensitivity as a function of the energy threshold for each NSB/hardware bin\footnote{The sensitivity is defined as the integral flux above an energy threshold giving $N_{\text{excess}} / \sqrt{N_{\text{bgd}}} = 5$, where $N_{\text{excess}}$ is the number of excess events and $N_{\text{bgd}}$ the number of background events, with additional constraints: $N_{\text{excess}} > 10$ and $N_{\text{excess}} > 0.05 N_{bgd}$.}. The obtained sensitivity curves are shown in figure \ref{fig:Sens}. To accumulate enough data in every NSB/hardware bin, we use data from a large zenith angle range going from 5$^\circ$ to 45$^\circ$. As the sensitivity and energy threshold depend strongly on the zenith angle and data sub-samples have different zenith angle distributions, the performances are corrected to correspond to the same reference zenith-angle distribution (average of all the data). To visualize the degradation caused by moonlight, the integral sensitivity computed for each NSB/hardware bin is divided by the one obtained under dark conditions at the same analysis-level energy threshold. 
The sensitivity degradation for nominal HV is constrained to be less than 10\% below 1\,TeV and all the curves are compatible within error bars above $\sim$300\,GeV. 
The only visible degradation is near the reconstruction-level energy threshold ($<$200\,GeV), where the sensitivity is 5-10\% worse. For Moon data taken with reduced HV, the sensitivity degradation lies between 15\% and 30\%.
This degradation is caused by a combination of a higher extracted-signal noise (see section 3 in \cite{MAGIC_moon}) and a smaller effective area. The degradation is even clearer in the UV-pass filter data, where the sensitivity is 60-80\% worse than the standard one. Such a degradation is expected, especially due to the fact that the filters reject more than 50\% of the Cherenkov light. Besides, sensitivity could also be affected by a poorer reconstruction of the images, especially in the pixels that are partially obscured by the filter frame ribs. At the highest energies ($>$2\,TeV) sensitivity seems to improve. This could be expected for bright images, that are less affected by noise, but higher statistics at those energies would be needed to
derive further conclusions.

\subsection*{Systematic uncertainties}

A study on the effect of the moonlight in the systematic uncertainties is performed in \cite{MAGIC_moon}. To summarize, those errors depend on the hardware configuration and the NSB level. For moderate moonlight (NSB~$<8 \times  \, \textit{NSB}_{\text{Dark}}$) observations with nominal HV, the additional systematic errors on the flux is below 10\%, raising the flux-normalization uncertainty (at a few hundred GeV) from 11\% \cite{upgrade2} to 15\%. For observations with reduced HV (NSB~$<18 \times  \, \textit{NSB}_{\text{Dark}}$) the additional systematic errors on the flux is $\sim$15\%, corresponding to a full flux-normalization uncertainty of 19\% after a quadratic addition. For UV-pass filter observations, the flux-normalization uncertainty increases to 30\%. The additional systematic on the reconstructed spectral index is negligible ($\pm$0.04) and the overall uncertainty is still $\pm$0.15 for all hardware/NSB configurations. The uncertainty of the energy scale is not affected by the moonlight. It may increase for reduced HV and UV-pass filter observations but this effect is included in the flux-normalization uncertainty increase
No additional systematic uncertainties have been found concerning the pointing accuracy.

Concerning day-to-day systematic errors, the additional uncertainties due to moonlight for observations with nominal HV (NSB~$<8 \times  \, \textit{NSB}_{\text{Dark}}$) are marginal and the overall day-to-day systematic errors are consistent with the 11\% in \cite{upgrade2}. For brighter moonlight that requires hardware modifications, the systematic errors get larger. The overall day-to-day systematic is estimated at $(15.4\pm3.2)$\% for reduced HV and $(13.2\pm3.4)$\% for UV-pass filters, corresponding to an additional systematic on top of the dark nominal HV systematic errors laying between 6\% and 18\%. For every hardware configuration, the additional day-to-day systematic errors is of the same order, or below, the systematic errors found for the overall flux.

\section{Conclusions}

For the first time the performance under moonlight of an IACT system is studied in detail with an analysis dedicated for such observations, including moonlight-adapted MC simulations. This study, presented in detail in \cite{MAGIC_moon}, includes data taken with three different hardware settings: nominal HV, reduced HV and UV-pass filters.

During moonlight, the additional noise results in a higher energy threshold increasing with the NSB level, which for zenith angles below 30$^{\circ}$ goes from $\sim$70\,GeV (at the reconstruction level) under dark conditions up to $\sim$300\,GeV in the brightest scenario studied (15-30~$\times  \, \textit{NSB}_{\text{Dark}}$). With a dedicated moonlight-adapted analysis, we are able to reconstruct the Crab Nebula spectrum in all the NSB/hardware bins considered. The flux obtained is compatible within 10\%, 15\% and 30\% with the one obtained under dark conditions for nominal HV, reduced HV and UV-pass filter observations, respectively.

An eventual degradation in the sensitivity is constrained to be below 10\% while observing with nominal HV under illumination levels $<8 \times  \, \textit{NSB}_{\text{Dark}}$. The sensitivity degrades by 15 to 30\% when observing with reduced HV and by 60 to 80\% when observing with UV-pass filters. No significant worsening on the angular resolution above 300\,GeV was observed.

The main benefit of operating the telescopes under moonlight is that duty cycle can be doubled, suppressing as well the need to stop observations around full Moon. The present study shows that, except for the energy threshold, the performance of IACT arrays is only moderately affected by moonlight. Hardware modifications to tolerate a strong sky brightness (reduced HV, UV-pass filters) seem to have more effect than the noise increase. The use of robust photodetectors, e.g. silicon photomultipliers, in the future should improve the performance under these bright conditions.

\end{document}